\documentclass[aps,pre,10pt,twocolumn,superscriptaddress]{revtex4-1}  
\usepackage{graphicx}  
\usepackage{dcolumn}   
\usepackage{bm}        
\usepackage{amssymb}   
\usepackage{physics}
\usepackage{graphicx,color}
\usepackage{multirow}
\usepackage{hyperref}
\usepackage{physics}
\usepackage{mathtools}

\def\ee{\mathrm e}
\def\OO{\mathcal O}

\begin{document}


\title{Comment on ``Kosterlitz-Thouless-type caging-uncaging transition in a quasi-one-dimensional hard disk system''}

\author{Yi Hu}
\email{yi.hu@duke.edu}
\affiliation{Department of Chemistry, Duke University, Durham, North Carolina 27708, USA}

\author{Patrick Charbonneau}
\affiliation{Department of Chemistry, Duke University, Durham, North Carolina 27708, USA}
\affiliation{Department of Physics, Duke University, Durham, North Carolina 27708, USA}

\date{\today}

\begin{abstract}
Huerta \emph{et al.} [Phys.~Rev.~Research 2, 033351 (2020)] report a power-law decay of positional order in numerical simulations of hard disks confined within hard parallel walls, which they interpret as a Kosterlitz-Thouless--type caging-uncaging transition. The proposed existence of such a transition in a quasi--one-dimensional (q1D) system, however, contradicts long-held physical expectations. To clarify if the proposed ordering persists in the thermodynamic limit, we introduce an exact transfer matrix approach to expeditiously generate equilibrium configurations for systems of arbitrary size. The power-law decay of positional order is found to extend only over finite distances. We conclude that the numerical simulation results reported are associated with a crossover, and not a proper thermodynamic phase transition.
\end{abstract}

\maketitle

\section{Introduction}
The work of Huerta \emph{et al.}~identifies a Kosterlitz-Thouless (KT)-type caging-uncaging transition in a system of hard disks confined between parallel walls~\cite{huerta2020kosterlitz}. That identification is based on the power-law decay of positional order in numerical simulations near close packing, and is supported by the detection of a narrow subpeak in the pair distribution function and of transverse excitation modes in the caging and uncaging regimes. 
The proposal is intriguing because the presence of a phase transition in such a system is physically unexpected. One-dimensional (1D) and quasi-one-dimensional (q1D) systems with short-ranged interactions have indeed long been considered incapable of exhibiting genuine phase transitions~\cite{van1950integrale,ruelle1999statistical,lieb2013mathematical}. 
(Singularities with respect to changes in structural quantities nonetheless remain possible~\cite{saryal2018multiple}.) 
KT transitions, however, differ from conventional phase transitions in many respects~\cite{berezinskii1972destruction,kosterlitz1973ordering}. 
They leave no thermal feature in the partition function and its derivatives, such as the specific heat~\cite{minnhagen1987two}, and are thus ``infinite-order'' in nature. 
The critical phase below the KT transition temperature also differs from conventional ordered phases in that it only exhibits quasi--long-range order. 
That critical phase is thus only identified from the power-law decay of spatial correlations, with a critical exponent value that changes with system conditions~\cite{kosterlitz1973ordering}. 
Do these features exempt KT transitions from traditional expectations for q1D systems? If not, how can one explain the power-law decay of the positional order observed in the simulations of Ref.~\cite{huerta2020kosterlitz}? 

These questions motivate our consideration of an exact transfer matrix scheme that provides equilibrium observables and correlation lengths in the thermodynamic limit, and thus sidesteps hurdles associated with thermalization and finite-size corrections in numerical simulations. 
Although the memory and computational complexities of transfer matrix treatments increase exponentially with the number of possible pair interactions~\cite{selke1988annni}, the approach  has already been demonstrated for q1D hard disks with up to next-nearest-neighbor interactions~\cite{godfrey2015understanding,hu2018correlation}. 
The model studied by Huerta \emph{et al.}~hence lies comfortably within the computationally accessible regime. 
However, because KT transitions leave no thermal signature and because the correlation length given by the transfer matrix is not directly related the decay of the longitudinal pair distribution function $g(x)$, standard transfer matrix schemes do not suffice. 
We thus introduce an approach for planting equilibrium configurations at minimal computational cost. By broadening the range of $g(x)$ compared to what Ref.~\cite{huerta2020kosterlitz} reports, we find that its power-law decay is truncated at large distances, and hence that the KT-like scaling observed in numerical simulations results from a smooth crossover rather than a genuine thermodynamic phase transition. 

\section{Planting Method}
Because the planting scheme proposed is generic and could be applied to any system solvable by transfer matrices, we first describe it in general terms, and then apply it to the specific q1D system of interest. 

\subsection{Transfer matrix setup}
Consider a system described by the Hamiltonian
\begin{equation}
\mathcal{H} = \sum_{i=1}^N V(a_i, a'_i),\end{equation}
for interactions between a unit $a_i$ and its subsequent unit, $a'_i = a_{i+1}$.  A unit could be, for instance, $m$ subsequent spins (in a lattice model) or particles (in an off-lattice model) with at most $m$-th nearest neighbor interactions. 

The key step consists of writing the partition function as the trace of a product of transfer matrices, such as $Z=\tr(\mathbf{K}^N)$ for $N$ identical units. For lattice models at inverse temperature $\beta$, matrix entries are 
\begin{equation}
K_{a a'} = \exp[-\beta V(a, a')],
\end{equation}
with $a$ and $a'$ indexing rows and columns, respectively, for the  $n$ possible states taken by $a$. For continuum-space (off-lattice) models, a discretization of space in $n$ segments of size $\delta a$ similarly gives 
\begin{equation} \label{eq:tmatcont}
K_{a a'} = \exp[-\beta V(a, a') ] \delta a,
\end{equation}
In both cases, the resulting $n \times n$ transfer matrix  can be used to write
\begin{equation}
Z = \tr( \mathbf{K}^N ) = \tr (\mathbf{U} \boldsymbol\Lambda \mathbf{U}^{-1}) = \lambda_0^N\left[ 1+\OO\left(\frac{\lambda_1}{\lambda_0}\right)^N \right]
\end{equation}
where $\mathbf{U} = (u_0, u_1, ...)$ is a matrix of eigenvectors. In the thermodynamic, $N \rightarrow \infty$, limit, the leading eigenvalue $\lambda_0^N$ asymptotically dominates the partition function, and hence the free energy per site is $f = -\log \lambda_0/\beta$. The $i$-th subleading eigenvalue can then also be used to obtain the $i$-th correlation length, $\xi_i = 1/\log(\lambda_0/|\lambda_i|)$. 

For 1D and q1D systems with finite-range interactions between units, the transfer matrix is finite with non-negative entries. According to the Perron-Frobenius theorem, the leading eigenvalue $\lambda_0$ is non-degenerate, i.e., $\lambda_0 > |\lambda_1|$, and the entries of leading left and right leading eigenvectors, $u_0$ and $u^{-1}_0$, are real and non-negative. It then follows that a system described by transfer matrices \emph{always} presents a finite correlation length for finite $\beta$ and thus cannot undergo a finite-temperature phase transition. This reasoning should also apply to finite-pressure KT-type transitions in q1D hard disks, as we consider below.

\subsection{Generating equilibrium configuration} \label{sec:eqconfig}
Unlike molecular simulations, which provide configurations in real space, transfer matrices are probabilistic objects. 
Structural observables commonly accessible in the former may thus not as easily be obtained from the latter. 
For instance, although the pair correlation, $g(x)$, can be computed by (inverse) Fourier transforming the structure factor computed as in Ref.~\cite{robinson2016glasslike}, this approach is not straightforwardly generalizable to many other structural observables.
To sidestep this difficulty equilibrium states can be \emph{planted} for subsequent analysis from the eigenvectors of these matrices. The marginal probability $P(a)$ (or $P(a) \delta a$ in off-lattice models) of a state $a$ in equilibrium configurations is indeed given by 
\begin{equation} \label{eq:mrgp}
P(a) = \frac{u_0^{-1}(a) u_0(a) }{\sum_i u_0^{-1}(i) u_0(i)},
\end{equation}
and the conditional probability that, given a state of $a$, the subsequent state is $a'$, $P(a'|a)$ (or $P(a'|a) \delta a'$ in off-lattice models) is
\begin{equation} \label{eq:cndp}
P(a'|a) = \frac{u_0^{-1}(a) K_{a a'} u_0(a') }{\sum_i u_0^{-1}(a) K_{a i} u_0(i)} = \frac{K_{a a'} u_0 (a') }{\lambda_0 u_0(a) }.
\end{equation}

Once the leading eigenvector of the transfer matrix is known, we can propagate an equilibrium configuration of arbitrary size by the following algorithm.
\begin{enumerate}
\item Generate a ``starting'' configuration according to Eq.~\eqref{eq:mrgp} by initializing a cumulative probability distribution array $Q(a)$ based on the marginal probability $P(a)$ such that
\begin{equation}
Q(a) = \sum_{i=1}^a u_0^{-1}(i) u_0(i) \sim \sum_{i=1}^a  P(i),
\end{equation}
for indexed states $a=1,...,n$, and $Q(0)=0$.
\item Generate a uniformly distributed random variable $\gamma\in [0, 1)$ and choose a state with index $a$ such that $Q(a-1) \le \gamma Q(n) < Q(a)$.
\item Initialize another cumulative probability distribution array $Q(a'|a)$ for the conditional probability $P(a'|a)$, such that
\begin{equation}
Q(a'|a) = \sum_{i=1}^{a'} K_{a i} u_0(i) \sim \sum_{i=1}^{a'}  P(i|a),
\end{equation}
for indexed states $a'=1,...,n$, and $Q(0|a)=0$.
\item Generate a random variable $\gamma$ again and choose the subsequent state $a'$ from $Q(a'-1|a) \le \gamma Q(n|a) < Q(a'|a)$.
\item Propagate subsequent states by setting $a \leftarrow a'$ and repeating steps 3 and 4.
\end{enumerate}

Because the state index $a$ denotes a continuous variable in off-lattice models, the sampling must then be interpolated. Specifically, $a$ and $a'$ are found by $a = f( \gamma Q(n) )$ and $a' = f( \gamma Q(n|a) | a)$, where $f(\cdot)$ and $f(\cdot|a)$ are interpolation functions that map $Q(a) \mapsto a$ and $Q(a'|a) \mapsto a'$, respectively. For sufficiently large $n$, the choice of interpolation method does not significantly affect the result, and for the q1D hard disk model considered a simple cubic interpolation scheme suffices.

\subsection{q1D hard disk scenario}
We now specialize to the case of hard disks of radius of $d$ in a channel infinite in the $x$ (longitudinal) direction, and defined by hard walls a distance $D$ apart in the $y$ (transverse) direction. The scaled transverse length free to disk centers is thus $h = (D-d)/d$. 
As in Ref.~\cite{huerta2020kosterlitz}, we further specialize to the case $D/d=3/2$. For $1 < D/d < 1+\sqrt{3}/2=1.866\ldots$, at most nearest-neighbor interactions between disks are geometrically possible, and thus $a_i $ in Eq.~\eqref{eq:tmatcont} is naturally taken as the vertical coordinates of disk centers, $y_i$. 
For such a q1D system, the partition function for the isothermal-isobaric (constant $NPT$) ensemble can be written using a transfer matrix approach~\cite{kofke1993hard,varga2011structural,gurin2015beyond}. (Although this scheme differs from the canonical, constant $NVT$ ensemble simulated in Ref.~\cite{huerta2020kosterlitz}, in the thermodynamic limit the two are rigorously equivalent.) An entry of the transfer matrix then reads
\begin{equation} \label{eq:tmatq1d}
K_{y y'} = \exp[-\beta F \sigma(y, y')]\sqrt{\delta y \delta y'} ,
\end{equation}
where $\sigma(y, y') = \sqrt{d^2 - (y-y')^2}$ is the contact distance between two neighboring disks and $F$ is the force associated with the longitudinal pressure.
Note that we have here replaced $\delta a$ in Eq.~\eqref{eq:tmatcont} with $\sqrt{\delta y \delta y'}$. The associated change of variable does not affect the eigenvalues of the transfer matrix, but keeps $\mathbf{K}$ symmetric even for uneven discretization of $y$, and hence $u = u^{-1}$. 

At high pressures, i.e., near close packing, disks are mostly confined near the walls. To improve the numerical accuracy of the discretization scheme, Ref.~\cite{godfrey2015understanding} suggested a change of variable $y \rightarrow t$, such that
\[ y(t) = a t + b \tanh (c\,t) \]
where $b=h/2$, $a=h/2 - b \tanh(c)$ and $c$ is adjustable. In particular, increasing $c$ refines the grid around the walls. The discretization of the new variable $t \in [-1, 1]$ is then uniformly spaced by $\delta t$ and the transfer matrix becomes
\begin{equation}
K_{t t'} = \exp[-\beta F \sigma(y(t), y(t')) \sqrt{\left.\dv{y}{t}\right|_t \cdot \left.\dv{y}{t}\right|_{t'}} \cdot \delta t],
\end{equation}
where $\dd y/\dd t = a+b\, c\, \mathrm{sech}^2(ct)$. Results for this regime are fairly insensitive to different choices of $c \sim \OO(1)$ and $n \ge 100$. Without loss of generality, we thus set $c=3$ and $n=1000$. Note that eigenvalue evaluation is then nearly instantaneous on a standard desktop computer.

The leading and subleading correlation lengths given by the transfer matrix for this systems correspond to the correlation length of the zig-zag order, $\xi_y=\xi_1 $, and of the longitudinal spacing, $\xi_{\delta x}=\xi_2$~\cite{godfrey2015understanding}. These lengths hence describe the decay at large distances, $|i-j| \rightarrow \infty$,  of
\begin{align}
g_y(|i-j|) &= \langle y_i y_j \rangle -\langle y_i\rangle^2\nonumber\\ 
&\sim (-1)^{|i-j|}\exp(-|i-j|/\xi_y) \label{eq:corzigzag}\\
g_{\delta x}(|i-j|) &= \langle \delta x_i \delta x_j \rangle - \langle \delta x_i \rangle^2  \nonumber\\ 
&\sim \exp(-|i-j|/\xi_{\delta x}),\label{eq:cordeltax}
\end{align}
respectively. Note that the oscillatory nature of Eq.~\eqref{eq:corzigzag} for zig-zag order is associated with $\lambda_2$ being negative~\cite{godfrey2015understanding}.

From the leading eigenvalues and eigenvectors of $\mathbf{K}$ we can also generate $y$-coordinates of disks according to the scheme described in Sec.~\ref{sec:eqconfig}. 
Longitudinal spacings between neighboring obstacles, $\delta x_i$, are then generated according to the distribution rule for the constant $NPT$ ensemble
\begin{equation}
P(\delta x) \sim \exp(-\beta F \delta x_i), \quad \delta x_i \ge \sigma(y_i, y_{i+1}).
\end{equation}
For each pressure considered, we generate configurations of longitudinal size $L \ge 500$, and compute the pair distribution function by averaging over 400 independent realizations.

\section{Results}
\begin{figure}[ht]
\includegraphics[width=0.95\columnwidth]{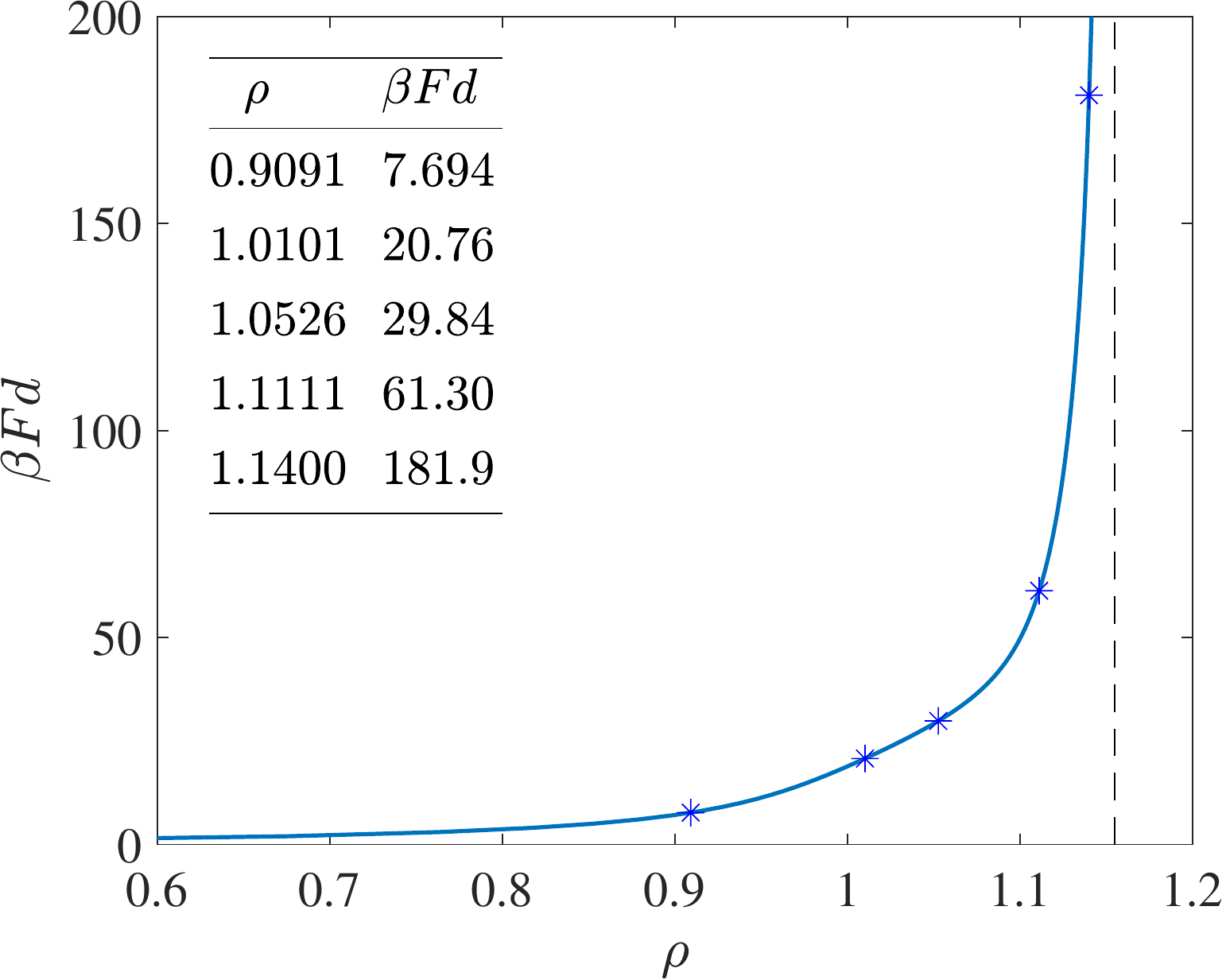}
\caption{Correspondence between the longitudinal force and density for $D/d=3/2$. Systems investigated in Ref.~\cite{huerta2020kosterlitz} and in this comment are marked by asterisks and are listed in the embedded table. The black dashed line denotes the close packing density $\rho = 2/\sqrt{3} = 1.1547\ldots$.
}
\label{fig:eos}
\end{figure}

\begin{figure*}[ht]
\includegraphics[width=0.98\textwidth]{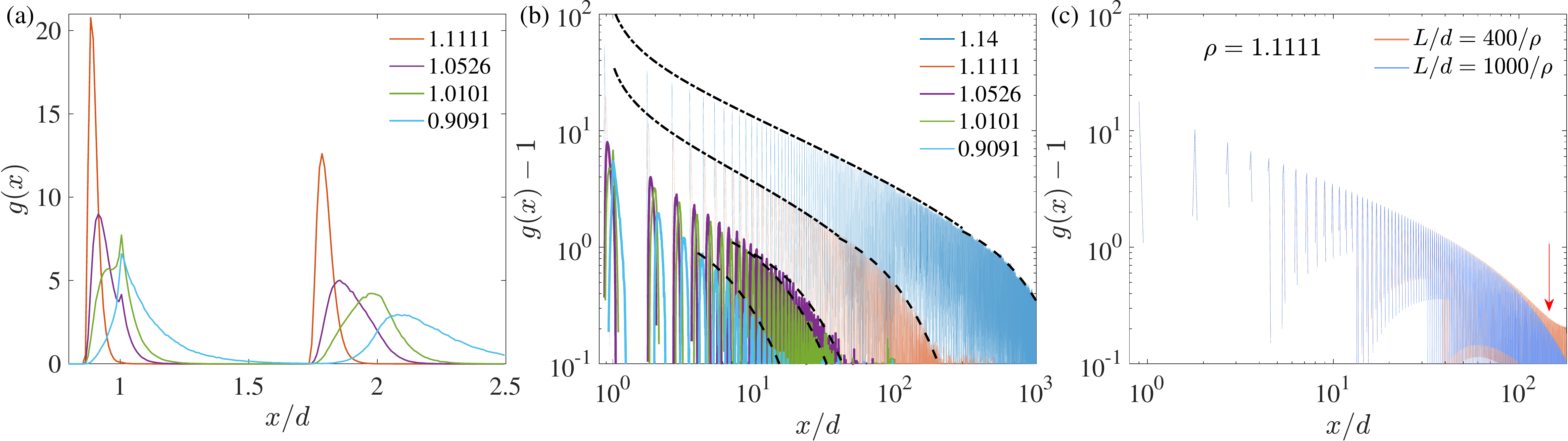}
\caption{Pair distribution function for $D/d=3/2$ and different densities at (a) short and (b) long distances obtained from planting, and (c) from the inverse DFT of the structure factor $S(q)$. In (a), the sub-peaks identified in Ref.~\cite{huerta2020kosterlitz} are clearly visible at short distances. In (b), the algebraic (power-law--like) decay, however, is clearly truncated at finite distances for all densities considered.
Peak heights are fitted to the decay form of Eq.~\eqref{eq:algdecay} (dash-dotted) and to an exponential (dashed) decay form at intermediate and long distances, respectively. In (c), an estimate of the finite-size correction in simulations is obtained by setting the effective system size $L$ in the spacing $\delta q = 2 \pi / L$ of the DFT, which gives $g(x) = g(L - x)$ as in periodic boundary condition. The excess of $g(x)$ at large $x$ for $L/d=400/\rho$ (arrow) qualitatively matches simulations results.}
\label{fig:gr}
\end{figure*}

\begin{figure}[ht]
\includegraphics[width=0.95\columnwidth]{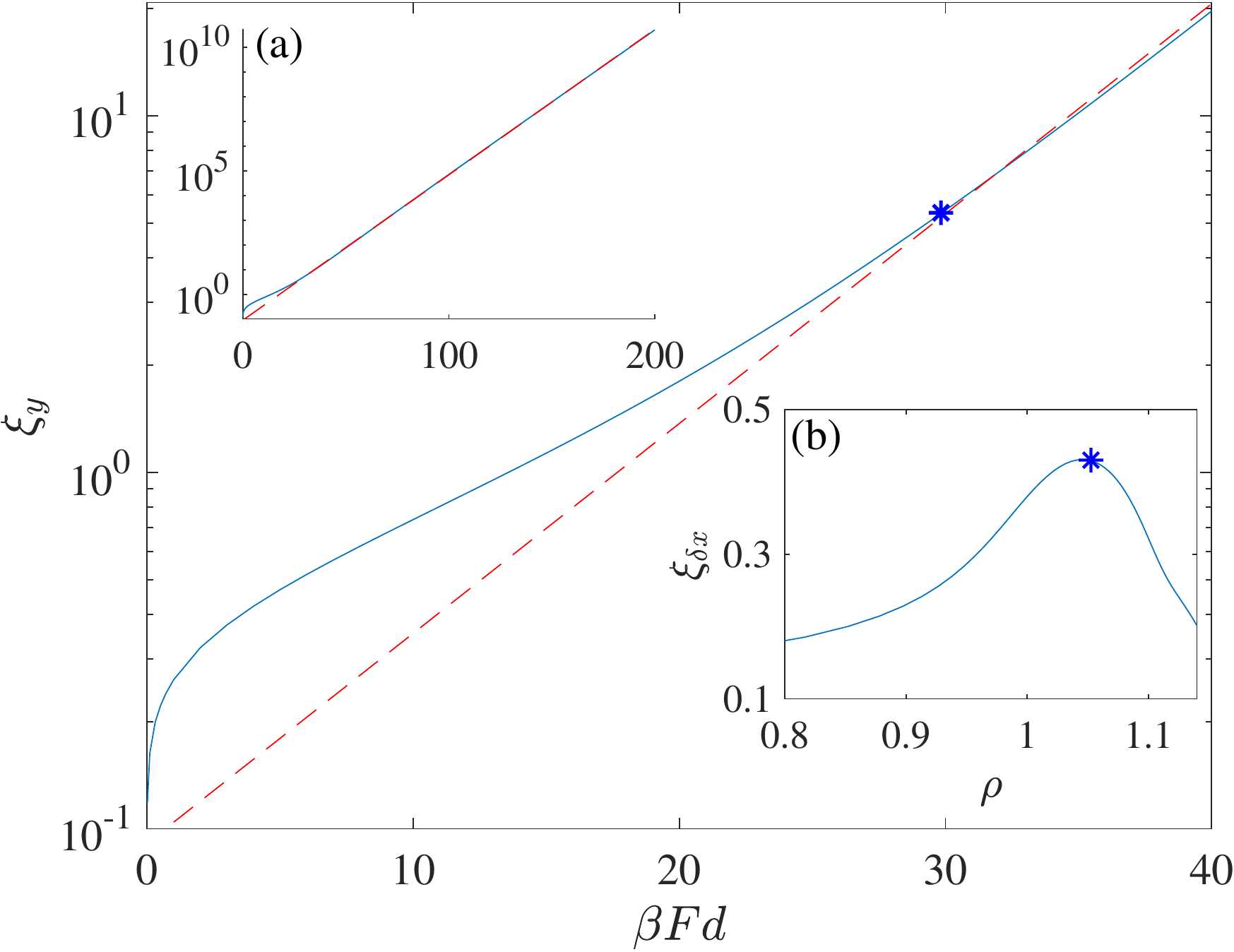}
\caption{Correlation length corresponding to zig-zag order, $\xi_y$, for $D/d=3/2$ (blue line). The transformation point $\rho_\mathrm{ost}=1.0526$ (asterisk) identified by Ref.~\cite{huerta2020kosterlitz} coincides with the onset of the exponential divergence (red dashed fitting line) of $\xi_y$ with pressure. (Inset a) Same quantity over a wider range of $\beta F$. (Inset b) Correlation length corresponding to the longitudinal spacing, $\xi_{\delta x}$, under the same conditions. The transformation point there coincides with the maximum of $\xi_{\delta x}$.}
\label{fig:clength}
\end{figure}

Because the transfer matrix is evaluated for the constant $NPT$ ensemble, we first determine the longitudinal force $F$ under the longitudinal number density $\rho = \lim_{N,L \rightarrow \infty} N/(L/d)$ (Fig.~\ref{fig:eos}), 
and then compute $g(x)$ from planted configurations (Fig.~\ref{fig:gr}).
At short distances, the transfer matrix and the simulation results are fully consistent, including the apparition of a small sub-peak at high densities (see \cite[Fig.~2(a)]{huerta2020kosterlitz}). 
At long distances, however, while Ref.~\cite{huerta2020kosterlitz} reports that $g(x) - 1$ decays with a characteristic power-law beyond a certain density, we find that for sufficiently large $x$, $g(x) - 1$ decays exponentially for all densities considered. 
For $\rho=1.1111$, in particular, we observe that the power-law--like decay terminates at $x \sim 100$, even though it persists at least up to $x \sim 200$ in simulations. Because the leading correlation length from the transfer matrix, $\xi_y = 364$ (Fig.~\ref{fig:clength}), is then very close to the simulated system size, $N=400$, this discrepancy is most likely a finite-size correction. The resulting self-interactions in $x$ through the periodic boundary condition obfuscates the exponential decay of $g(x)-1$. As additional evidence, we note that $g(x)$ computed from the inverse discrete Fourier transform (DFT) of the structure factor $S(q)$ (obtained as in Ref.~\cite{robinson2016glasslike}) can be made to look like the simulation results by choosing a discretization spacing $\delta q$ that corresponds to a finite system size $L = 2\pi/\delta q$ (Fig.~\ref{fig:gr}(c)).  

Is the intermediate-range algebraic decay then an echo of a two-dimensional transition? The consideration of a purely 1D model suggests not.
Recall that for 1D rigid rods of length $d$~\cite{frenkel1946kiinetic,salsburg1953molecular},
\begin{equation} \label{eq:exactgr}
g(\hat{x}) = (1+\hat{p} ) \sum_{k=1}^{\lfloor \hat{x} \rfloor} \frac{\ee^{-\hat{p}(\hat{x}-k)} \left[\hat{p}(\hat{x}-k)\right]^{k-1} }{(k-1)!},
\end{equation}
where $\hat{x} = x/d$ and $\hat{p} = \beta F d$ are the reduced distance and pressure, respectively. Note that at high $\hat{p}$ and small $\hat{x}$, the height of the $k$-th peak is approximately the maximal value of the $k$-th summand,
\begin{align}
\hat{x}_\mathrm{max}^k &= k  + (k-1)/\hat{p}, \\
g(\hat{x}_\mathrm{max}^k) &= (1+\hat{p}) \frac{[(k-1)/\ee]^{k-1}}{(k-1)!}\nonumber \\
&= \frac{1+\hat{p}}{\sqrt{2 \pi (k-1)}} + \OO [ (\frac{1}{k-1})^{\frac{3}{2}} ].
\end{align}
Note also that at high pressures disks in the q1D system studied can be viewed as 1D hard rods of effective length $d'= d \sqrt{1-h^2} = d\cdot\sqrt{3}/2$. Because the variance of $y$ is non-zero, however, the peak height of $g(x)$ is also reduced by a constant multiplicative factor, $c<1$. By replacing $\hat{p} = \beta F d'$ and $k = \frac{x_\mathrm{max}/d'+ 1/\hat{p}}{1+1/\hat{p}}$, we thus obtain an approximate form for the algebraic $x_\mathrm{max}^{-1/2}$ decay of the peak height,
\begin{equation} \label{eq:algdecay}
 g(x_\mathrm{max} ) - 1 = c \cdot \frac{1+ \hat{p} }{\sqrt{2 \pi (k-1)}} - 1.
\end{equation}
Setting $c=0.7$ nicely fits both $\rho=1.1111$ and $1.1400$ for $g(x_\mathrm{max})>\OO(1)$ and $x_\mathrm{max} > \OO(1)$ (Fig.~\ref{fig:gr}(b)). 
For $g(x_\mathrm{max})-1 \lesssim \OO(1)$, however, the single summand approximation to $g(x)$ breaks down. The algebraic decay thus only persist up to $\sqrt{k} \sim \hat{p}$, i.e., $x/d \sim (\beta Fd)^2$. Interestingly, the exponential decay of $g(x)-1$ as $x \rightarrow \infty$, which truncates the algebraic decay, is also a 1D feature~\cite{perry1972decay}. 
This decay is controlled by the characteristic length $\xi_{g(x)}$ of a Tonk gas~\cite[Eq. 24]{hu2018correlation}. As $h \rightarrow 0$, the rows of the transfer matrix converge, and hence the associated correlation lengths vanish, but $\xi_{g(x)}$ persists. In other words, as long as $h$ is small, the algebraic decay reported in Ref.~\cite{huerta2020kosterlitz} is essentially a 1D feature, and thus not an echo of KT-type transition. Beyond the nearest-neighbor interaction regime, $h > \sqrt{3}/2$, however, the decay of $g(x)$ does genuinely become quite rich~\cite{fu2017assembly}.

Finally, Ref.~\cite{huerta2020kosterlitz} investigates the distribution of longitudinal spacing $\delta x$, over which window-like defects become substantial, and identified the onset of caging-uncaging transformation around $\rho_\mathrm{ost} = 1.0526$. 
Interestingly, this signature can also be found in $\xi_y$ and $\xi_{\delta x}$. 
These correlation lengths, which have been previously studied for q1D systems~\cite{varga2011structural,godfrey2015understanding,hu2018correlation}--including for $D/d=3/2$--are shown in Fig.~\ref{fig:clength} for reference. 
The reported $\rho_\mathrm{ost}$ coincides with the onset of exponential growth of $\xi_y$ with pressure (longitudinal force), as analyzed in Ref.~\cite{varga2011structural}. 
It further coincides with the maximum of $\xi_{\delta x}$ reported in Ref.~\cite{hu2018correlation}. We thus conclude that the phenomenon reported in Ref.~\cite{huerta2020kosterlitz} is also associated with that crossover.

\section{Summary}

Based on direct evidence gleaned from an exact planting scheme derived from transfer matrices, we have demonstrated that the power-law--like decay of positional order reported Ref.~\cite{huerta2020kosterlitz} only persists over finite distances. The pre-asymptotic power-law decay of the pair correlation is rooted in 1D physics, and not in any KT-type transition or its echo. The suggested uncaging transformation in Ref.~\cite{huerta2020kosterlitz} instead coincides with anomalies identified in earlier studies, and may thus be considered as a part of that same crossover.

\begin{acknowledgments}
We thank Michael~A.~Moore for stimulating discussions. We acknowledge support from the Simons Foundation (\#454937) and from the  National Science Foundation Grant No. DMR-1749374.
Data relevant to this work have been archived and can be accessed at the Duke Digital Repository.
\end{acknowledgments}

\bibliography{abbrev}

\end{document}